\documentclass[english,10pt, aps, prd, a4paper, preprintnumbers,floatfix, nofootinbib, showpacs, superscriptaddress, notitlepage]{revtex4-1}
\usepackage{graphicx}
\usepackage{rotate}
\usepackage[utf8]{inputenc}
\usepackage{amsmath}
\usepackage{amssymb}
\usepackage{float}
\usepackage{comment}
\usepackage{soul}
\usepackage[usenames,dvipsnames]{color}
\usepackage[colorinlistoftodos]{todonotes}
\usepackage{array}
\usepackage{amsmath}
\usepackage{amssymb}
\usepackage[colorlinks=true,citecolor=darkred,urlcolor=darkred, pdfborder={0 0 0}]{hyperref}
\usepackage[normalem]{ulem}
\pdfoutput=1
\newcommand {\ignore}[1]{}
\definecolor{darkred}{rgb}{0.6,0,0}

\def\tt1{$\mathrm{SU(3) \otimes SU(3)_L \otimes U(1)}$ }
\def\3311{$\mathrm{SU(3) \otimes SU(3)_L \otimes U(1)_X \otimes U(1)_{N}}$ }
\def\0331{$\mathrm{SU(3) \otimes SU(3)_L \otimes U(1)_X }$ }
\def\gsim{\raise0.3ex\hbox{$\;>$\kern-0.75em\raise-1.1ex\hbox{$\sim\;$}}}
\def\lsim{\raise0.3ex\hbox{$\;<$\kern-0.75em\raise-1.1ex\hbox{$\sim\;$}}}

\newcommand{\sm}{{Standard Model }}
\definecolor{mightnightblue}{RGB}{25,25,112}
\definecolor{brown}{rgb}{0.59, 0.29, 0.0}

\def\21{$\mathrm{SU(2)_L \otimes U(1)_Y}$}

\def\sm{standard model }

\newcommand{\Antonio}[1]{{\color{cyan}#1}}

\newcommand{\AddrAHEP}{  AHEP Group, Institut de F\'{i}sica Corpuscular --
  CSIC/Universitat de Val\`{e}ncia, Parc Cient\'ific de Paterna.\\
 C/ Catedr\'atico Jos\'e Beltr\'an, 2 E-46980 Paterna (Valencia) - SPAIN}
\input{tcilatex}
\begin{document}

\title{\color{BrickRed} Simple theory for scotogenic dark matter
with residual matter-parity}
\author{A. E. C\'{a}rcamo Hern\'{a}ndez}
\email{antonio.carcamo@usm.cl}
\affiliation{Departamento de F\'isica, Universidad T\'{e}cnica Federico Santa Mar\'{\i}a,\\
Casilla 110-V, Valpara\'{\i}so, Chile}
\author{Jos\'{e} W. F. Valle}
\email{valle@ific.uv.es}
\affiliation{\AddrAHEP}
\author{Carlos A. Vaquera-Araujo}
\email{vaquera@fisica.ugto.mx}
\affiliation{Departamento de F\'isica, DCI, Campus Le\'on, Universidad de
Guanajuato, Loma del Bosque 103, Lomas del Campestre C.P. 37150, Le\'on, Guanajuato, M\'exico}
\affiliation{Consejo Nacional de Ciencia y Tecnolog\'ia, Avenida Insurgentes Sur
1582. Colonia Cr\'edito Constructor, Alcald\'ia Benito Ju\'arez, C.P. 03940,
Ciudad de M\'exico, M\'exico}
\date{\today }





\begin{abstract}
\vspace{1cm}

Dark matter stability can result from a residual matter-parity symmetry surviving spontaneous breaking of an extended gauge symmetry.
We propose the simplest scotogenic dark matter completion of the original SVS theory~\cite{Singer:1980sw}, in which the ``dark sector'' particles as well as matter-parity find a natural
theoretical origin within the model. We briefly comment on its main features. 

\end{abstract}

\maketitle

\section{Introduction}

\label{sec:introduction}

The nature of dark matter remains mysterious, though a lot of progress has
been made on what dark matter should \textit{not} be~\cite{Bertone:2004pz}.
Many particle dark matter candidates have been proposed in agreement with
astrophysical and cosmological observations, in particular the so-called
Weakly Interacting Massive Particles, or WIMPs, have attracted a lot of
attention. From a theory point of view it would be desirable that the
particle dark matter candidate should obey two requirements:

\begin{enumerate}
\item fit in a broder scheme accounting for other shortcomings of the
standard model,

\item have its stability on cosmological scales naturally protected by a
symmetry.
\end{enumerate}

The existence of supersymmetry would provide a WIMP candidate, the Lightest
supersymmetric particle, though it fails to obey the above requirements,
since its stability is assumed as a result of R-parity conservation, an 
\textit{ad hoc} symmetry~\cite{Jungman:1995df}. Moreover, the LSP does not
relate to other problems of the standard model except, possibly, the
technical aspects associated to the hierarchy problem.

Neutrino mass generation is one of the basic open challenges in particle physics and it could well be that it may be directly related to the understanding of dark matter.
Indeed, WIMP dark matter could mediate neutrino mass generation~\cite{Ma:2006km}.
This idea, realized within the simplest standard model gauge structure, has been studied in many papers over the past few 
years~\cite{Hirsch:2013ola,Hernandez:2015hrt,Merle:2016scw,Bernal:2017xat,Rojas:2018wym,Restrepo:2019ilz,Avila:2019hhv,Arbelaez:2019ofg}.

When the gauge symmetry is extended, it can happen that there is a ``dark symmetry'' called matter-parity, that remains conserved after spontaneous symmetry breaking.
In this case the lightest odd-particle will be automatically stable and can play the role of dark matter.
Indeed, this has been shown to be the case in the context of the \3311 electroweak extension of the standard model~\cite{Alves:2016fqe,Kang:2019sab,Leite:2019grf,VanLoi:2019eax}.

Here we construct a non-supersymmetric scenario for scotogenic dark matter
in which dark matter stability results naturally from the residual matter-parity symmetry.
The construction provides the simplest dark matter completion of the original SVS theory~\cite{Singer:1980sw} by incorporating ``automatically'' a stable scotogenic dark matter candidate.

The theory is minimal, as it uses only particles already present in the original SVS theory to make up the ``dark'' sector,
with the residual matter-parity resulting from the extended symmetry breaking dynamics~\footnote{A singlet scalar is also added to break the degeneracy of the neutral scalars,
  needed to close the scotogenic neutrino mass loop.}.
This way it provides an elegant origin for the scalar dark doublet introduced \textit{ad hoc} in other dark matter constructions, of the Inert Higgs Doublet type~\cite%
{Deshpande:1977rw,LopezHonorez:2006gr,Dolle:2009fn,Honorez:2010re,LopezHonorez:2010tb}.
The latter is naturally identified here with the inert electroweak doublet contained in one of the triplet Higgs scalars required to ensure adequate breaking of the
extended $\mathrm{SU(3)_L}$ gauge symmetry. If lightest, its stability becomes automatic because of the residual matter-parity gauge symmetry.

The paper is organized as follows: in Sec.\ref{sec:model} we
sketch the theory setup and quantum numbers. In Sec.~\ref{sec:Scalars} we summarize the
scalar sector and in Sec.~\ref{sec:Yukawa} we
describe the Yukawa couplings and the neutrino mass generation mechanism.  Sec.~\ref{sec:DM} is devoted to the analysis of the dark matter candidates of the model. Finally in Sec.~\ref{sec:Discussion} we present a short discussion
and conclude.

\section{The model}

\label{sec:model}

Our starting point is a variant of the model introduced in \cite{Alves:2016fqe} based on the \3311 gauge symmetry. The main motivation for the extra U(1)$_N$ is
to allow for a fully gauged $B-L$ symmetry \cite{Dong:2014wsa,Dong:2015yra}.
In our model, electric charge and $B-L$ are embedded into the gauge symmetry
as 
\begin{align}
Q & = T_3-\frac{T_8}{\sqrt{3}}+X, \\
B-L & = -\frac{2}{\sqrt{3}}T_8+N,
\end{align}
with $T_i$ $(i = 1,2,3,...,8)$, $X$ and $N$ as the respective generators of
SU(3)$_L$, U(1)$_X$ and U(1)$_N$.

In the present model, after spontaneous symmetry breaking a residual
discrete symmetry arises as a remnant from the $B-L$ symmetry breakdown. Its
role is analogous to that of $R$-parity in supersymmetric theories, we call
it matter-parity, $M_P=(-1)^{3(B-L)+2s}$. The stability of the lightest $M_P$%
-odd particle leads to a potentially viable WIMP dark matter candidate.

The particle content of the model is shown in Table \Antonio{\ref{tab:tab3311}}. Here,
left-handed leptons $l_{aL}$; $a =1,2,3$ transform as triplets under $%
\mathrm{SU(3)_L }$, 
\begin{equation}
l_{aL}=%
\begin{pmatrix}
\nu_a \\ 
e_a \\ 
N_a%
\end{pmatrix}%
_L,
\end{equation}
and the third component is precisely the  $M_P$-odd singlet fermion needed in a scotogenic neutrino mass generation mechanism. Anomaly cancellation
requires that two generations of quarks $q_{i L}$; $i = 1,2$ must transform
as anti-triplets and one as a triplet~\cite{Singer:1980sw}~\footnote{Here we follow mainly the gauged B-L extension of the original reference. Many other works exist, see also~\cite{Valle:1983dk,Pisano:1991ee,Frampton:1992wt,Hoang:1995vq,Boucenna:2014ela,Okada:2015bxa,Fonseca:2016tbn,CarcamoHernandez:2017cwi}.}, 
\begin{equation}
q_{i L}=%
\begin{pmatrix}
d_i \\ 
-u_i \\ 
D_i%
\end{pmatrix}%
_L\qquad q_{3L}=%
\begin{pmatrix}
u_3 \\ 
d_3 \\ 
U_3%
\end{pmatrix}%
_L,
\end{equation}
This choice predicts three generations of quarks and leptons (the same as
the number of colors), an important feature of this class of models. 
\begin{table}[ht]
\centering
\begin{tabular}{|c|c|c|c|c|c|c|}
\hline
\hspace{0.2cm} Field \hspace{0.2cm} & \hspace{0.2cm}SU(3)$_c$ \hspace{0.2cm}
& \hspace{0.2cm} SU(3)$_L$ \hspace{0.2cm} & \hspace{0.2cm}U(1)$_X$ \hspace{%
0.2cm} & \hspace{0.2cm}U(1)$_N$ \hspace{0.2cm} & \hspace{0.2cm} $Q$\hspace{%
0.2cm} & \hspace{0.2cm} $M_P = (-1)^{3(B-L)+2s}$ \hspace{0.2cm} \\ 
\hline\hline
$q_{i L}$ & \textbf{3} & $\overline{{\mathbf{3}}} $ & 0 & 0 & $(-\frac{1}{3},%
\frac{2}{3},-\frac{1}{3})^T$ & $(++-)^T$ \\ 
$q_{3L}$ & \textbf{3} & \textbf{3} & $\frac{1}{3}$ & $\frac{2}{3}$ & $(\frac{%
2}{3},-\frac{1}{3},\frac{2}{3})^T$ & $(++-)^T$ \\ 
$u_{aR}$ & \textbf{3} & \textbf{1} & $\frac{2}{3}$ & $\frac{1}{3}$ & $\frac{2%
}{3}$ & $+$ \\ 
$d_{aR}$ & \textbf{3} & \textbf{1} & $-\frac{1}{3}$ & $\frac{1}{3}$ & $-%
\frac{1}{3}$ & $+$ \\ 
$U_{3R}$ & \textbf{3} & \textbf{1} & $\frac{2}{3}$ & $\frac{4}{3}$ & $\frac{2%
}{3}$ & $-$ \\ 
$D_{i R}$ & \textbf{3} & \textbf{1} & $-\frac{1}{3}$ & $-\frac{2}{3}$ & $-%
\frac{1}{3}$ & $-$ \\ 
$l_{aL}$ & \textbf{1} & \textbf{3} & $-\frac{1}{3}$ & $-\frac{2}{3}$ & $%
(0,-1,0)^T$ & $(++-)^T$ \\ 
$e_{a R}$ & \textbf{1} & \textbf{1} & $-1$ & $-1$ & $-1$ & $+$ \\ \hline
$\nu_{i R}$ & \textbf{1} & \textbf{1} & $0$ & $-4$ & $0$ & $-$ \\ 
$\nu_{3R}$ & \textbf{1} & \textbf{1} & $0$ & $5$ & $0$ & $+$ \\ 
$N_{aR}$ & \textbf{1} & \textbf{1} & $0$ & $0$ & $0$ & $-$ \\ \hline\hline
$\eta$ & \textbf{1} & \textbf{3} & $-\frac{1}{3}$ & $\frac{1}{3}$ & $%
(0,-1,0)^T$ & $(++-)^T$ \\ 
$\rho$ & \textbf{1} & \textbf{3} & $\frac{2}{3}$ & $\frac{1}{3}$ & $%
(1,0,1)^T $ & $(++-)^T$ \\ 
$\chi$ & \textbf{1} & \textbf{3} & $-\frac{1}{3}$ & $-\frac{2}{3}$ & $%
(0,-1,0)^T$ & $(--+)^T$ \\ 
$\phi$ & \textbf{1} & \textbf{1} & $0$ & $2$ & $0$ & $+$ \\ \hline
$\sigma$ & \textbf{1} & \textbf{1} & $0$ & $1$ & $0$ & $-$ \\ \hline
\end{tabular}%
\caption{3311 model particle content ($a=1,2,3$ and $i =1,2$ represent
generation indices). Note the non-standard charges of ``right handed
neutrinos'' $\protect\nu_R$.}
\label{tab:tab3311}
\end{table}

Besides the fields contained in \cite{Alves:2016fqe}, the model contains only one
scalar singlet $\sigma$. This field will play an important role in the
neutrino mass generation mechanism by breaking the degeneracy of the real
and imaginary parts of the scalar exchanged in the scotogenic loop. The original scotogenic proposal includes a dark $\mathrm{SU(2)_L}$ doublet.
  A key observation of the present work is that such dark doublet is already present in the original SVS model when promoted to a \3311 gauge symmetry,
  it is naturally identified with the first two components of the $\chi$ triplet, that are already $M_P$-odd.

Notice the unconventional $\mathrm{U(1)_N}$ charges of the $\nu_R$ fields. Due to this
choice the tree level neutrino mass is absent. The two $\nu_{iR}$ neutrinos
can acquire a majorana mass after spontaneous symmetry breaking (SSB) by the inclusion of a scalar field
transforming as $( \mathbf{1} , \mathbf{1},0,8)$ and $\nu_{3R}$ requires a
scalar with quantum numbers $( \mathbf{1} , \mathbf{1},0,-10)$. In the
present work we do not include those fields in order to keep the analysis of
the scalar sector as simple as possible.

The gauged $B-L$ symmetry is spontaneously broken by two units as the
singlet scalar $\phi$ develops a vacuum expectation value (VEV), leaving a
discrete remnant symmetry $M_P=(-1)^{3(B-L)+2s}$. The most general VEV
alignment for the scalar fields compatible with the preservation of $M_P$
symmetry is 
\begin{equation}
\langle\eta\rangle=\frac{1}{\sqrt{2}}(v_1,0,0)^T,\quad \langle\rho\rangle=%
\frac{1}{\sqrt{2}}(0,v_2,0)^T, \quad\langle\chi\rangle=(0,0,w)^T, \quad
\langle\phi\rangle=\frac{1}{\sqrt{2}}\Lambda,\quad \langle\sigma\rangle=0.
\end{equation}
In this work we will assume the hierarchy $w, \Lambda,\gg v_1, v_2 $, such
that the SSB pattern of the model is 
\begin{align}
SU(3)_{C}^{}\times &SU(3)_{L}^{}\times U(1)_{X}^{}\times U(1)_{N}^{}  \notag
\\
&\downarrow w, \Lambda  \notag \\
SU(3)_C^{} &\times SU(2)_{L}^{}\times U(1)_{Y}^{}\times M_P^{}  \notag \\
&\downarrow v_1,v_2  \notag \\
SU(3)_C^{} & \times U(1)_{Q}^{}\times M_P^{}\,.
\end{align}
\begin{figure}[!h]
\includegraphics[width = .7\textwidth]{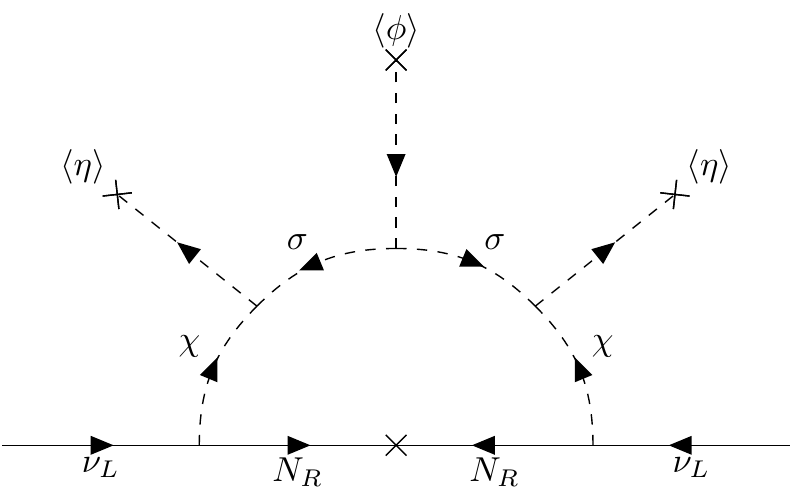}
\caption{ Feynman-loop diagram contributing to the light active Majorana
  neutrino mass matrix.}
\label{Scotogenicdiagram}
\end{figure}
Tree level light active neutrino masses
are forbidden in this basic setup. Small masses for the light active neutrinos are only generated at one loop level via a radiative seesaw mechanism mediated by the
CP-even and CP-odd parts of the first component of the $\mathrm{SU(3)_L}$ scalar triplet $\chi$ as well as by the gauge singlet right handed Majorana
neutrinos, as shown in Figure \ref{Scotogenicdiagram}.

\section{Scalars}

\label{sec:Scalars}

In this section we discuss the scalar sector of our model. The scalar multiplets are decomposed as follows,
\begin{equation}
\label{scalar3plets3}
\eta=\left(
\begin{array}{c}
\frac{v_1+s_1+i a_1}{\sqrt{2}}\\
\eta_2^{-}\\
\frac{s'_3+i a'_3}{\sqrt{2}}
\end{array}
\right),\quad
\rho=\left(
\begin{array}{c}
\rho_1^{+}\\
\frac{v_2+s_2+i a_2}{\sqrt{2}}\\
\rho_3^{+}
\end{array}
\right),\quad
\chi=\left(
\begin{array}{c}
\frac{s'_1+i a'_1}{\sqrt{2}} \\
 \chi_2^{-} \\
 \frac{w+s_3+i a_3}{\sqrt{2}} \\
\end{array}
\right),\quad
\phi=\frac{\Lambda+s_{\phi}+i a_{\phi}}{\sqrt{2}},\quad
\sigma=\frac{s_{\sigma}+i a_{\sigma}}{\sqrt{2}}.
\end{equation}

The scalar potential invariant under the symmetries of the model takes the form: 
\begin{equation}
\begin{split}
V=& \mu _{1}^{2}\rho ^{\dagger }\rho +\mu _{2}^{2}\chi ^{\dagger }\chi +\mu
_{3}^{2}\eta ^{\dagger }\eta +\mu _{4}^{2}\phi ^{\dagger }\phi +\mu
_{5}^{2}\sigma ^{\dagger }\sigma \\
& +\lambda _{1}(\rho ^{\dagger }\rho )^{2}+\lambda _{2}(\chi ^{\dagger }\chi
)^{2}+\lambda _{3}(\eta ^{\dagger }\eta )^{2} \\
& +\lambda _{4}(\rho ^{\dagger }\rho )(\chi ^{\dagger }\chi )+\lambda
_{5}(\rho ^{\dagger }\rho )(\eta ^{\dagger }\eta )+\lambda _{6}(\chi
^{\dagger }\chi )(\eta ^{\dagger }\eta ) \\
& +\lambda _{7}(\rho ^{\dagger }\chi )(\chi ^{\dagger }\rho )+\lambda
_{8}(\rho ^{\dagger }\eta )(\eta ^{\dagger }\rho )+\lambda _{9}(\chi
^{\dagger }\eta )(\eta ^{\dagger }\chi ) \\
& +\lambda _{10}(\phi ^{\dagger }\phi )(\rho ^{\dagger }\rho )+\lambda
_{11}(\phi ^{\dagger }\phi )(\chi ^{\dagger }\chi )+\lambda _{12}(\phi
^{\dagger }\phi )(\eta ^{\dagger }\eta ) \\
& +\lambda _{13}(\sigma ^{\dagger }\sigma )(\rho ^{\dagger }\rho )+\lambda
_{14}(\sigma ^{\dagger }\sigma )(\chi ^{\dagger }\chi )+\lambda _{15}(\sigma
^{\dagger }\sigma )(\eta ^{\dagger }\eta ) \\
& +\lambda _{16}(\phi ^{\dagger }\phi )^{2}+\lambda _{17}(\sigma ^{\dagger
}\sigma )^{2}+\lambda _{18}(\phi ^{\dagger }\phi )(\sigma ^{\dagger }\sigma
)+\lambda _{19}\left[ (\sigma ^{\dagger }\phi )(\eta ^{\dagger }\chi )+\text{%
h.c.}\right] \\
& +\frac{1}{\sqrt{2}}\left[- \mu_{t}\rho \eta \chi +\mu_{s}\phi ^{\dagger
}\sigma \sigma +\mu_{u}(\eta ^{\dagger }\chi )\sigma +\text{h.c.}%
\right] \ ,
\end{split}%
\end{equation}
where the $\lambda _{k}$ ($k=1,2,\cdots ,19$) are dimensionless parameters
whereas the $\mu _{r}$ ($r=1,2,\cdots ,5$), $\mu _{t}$, $\mu _{s}$, $\mu
_{u} $ are dimensionful parameters. Note that, to ensure
   the preservation of $M_P$, we assume $\mu _{5}^{2}>0$. The minimization
condition of the scalar potential yields the following relations: 
\begin{eqnarray}
\mu _{1}^{2} &=&\frac{v_1 w \mu _t-v_2 \left(\lambda _{10} \Lambda ^2+2 \lambda _1 v_2^2+\lambda _5 v_1^2+\lambda _4 w^2\right)}{2 v_2} ,  \notag \\
\mu _{2}^{2} &=&\frac{v_1 v_2 \mu _t-w \left(\lambda _{11} \Lambda ^2+\lambda _4 v_2^2+\lambda _6 v_1^2+2 \lambda _2 w^2\right)}{2 w},  \notag \\
\mu _{3}^{2} &=&\frac{v_2 w \mu _t-v_1 \left(\lambda _{12} \Lambda ^2+2 \lambda _3 v_1^2+\lambda _5 v_2^2+\lambda _6 w^2\right)}{2 v_1},  \notag \\
\mu _{4}^{2} &=&-\frac{1}{2} \left(2 \lambda _{16} \Lambda ^2+\lambda _{10} v_2^2+\lambda _{12} v_1^2+\lambda _{11} w^2\right).
\end{eqnarray}%
From the analysis of the scalar potential, we find that the squared mass
matrix for charged scalars, in the basis $\left( \eta _{2}^{+},\rho
_{1}^{+},\chi _{2}^{+},\rho _{3}^{+}\right) $ versus $\left( \eta
_{2}^{-},\rho _{1}^{-},\chi _{2}^{-},\rho _{3}^{-}\right)$, takes the form:
\begin{eqnarray}
M_{C} &=&\left( 
\begin{array}{cc}
M_{C}^{\left( 1\right) } & 0_{2\times 2} \\ 
0_{2\times 2} & M_{C}^{\left( 2\right) }%
\end{array}%
\right) , \\
M_{C}^{\left( 1\right) } &=&\frac{1}{2} \left(v_1 v_2 \lambda _8+w \mu _t\right)\left(
\begin{array}{cc}
 \frac{v_2}{v_1} & 1 \\
 1 & \frac{v_1}{v_2} \\
\end{array}
\right), \\
M_{C}^{\left( 2\right) } &=&\frac{1}{2}\left(w v_2 \lambda _7+v_1 \mu _t\right)\left(
\begin{array}{cc}
 \frac{v_2 }{w} & 1 \\
 1& \frac{w}{ v_2} \\
\end{array}
\right) .
\end{eqnarray}%
Note that the squared mass matrix $M_{C}$ has two vanishing eigenvalues
which correspond to the Goldstone bosons 
\begin{equation}
G^{\pm}=\frac{v_1\eta_2^{\pm}-v_2\rho_1^{\pm}}{\sqrt{v_1^2+v_2^2}},\qquad G'^{\pm}=\frac{w\chi_2^{\pm}-v_2\rho_3^{\pm}}{\sqrt{w^2+v_2^2}}
\end{equation}
associated to the longitudinal components of the $W^{\pm }$ and $W^{\prime\pm }$.
The massive eigenstates are the physical charged scalar bosons $H_{1}^{\pm }$ and $H_{2}^{\pm }$
\begin{equation}
\begin{split}
H_{1}^{\pm }&=\frac{v_2\eta_2^{\pm}+v_1\rho_1^{\pm}}{\sqrt{v_1^2+v_2^2}},
\qquad m^2_{H_{1}^{\pm }}=\frac{\left(v_1^2+v_2^2\right) \left(w \mu _t+\lambda _8 v_1 v_2\right)}{2 v_1 v_2},\\
H_{2}^{\pm }&=\frac{v_2\chi_2^{\pm}+w\rho_3^{\pm}}{\sqrt{w^2+v_2^2}},\qquad m^2_{H_{2}^{\pm }}=\frac{\left(v_2^2+w^2\right) \left(v_1 \mu _t+\lambda _7 v_2 w\right)}{2 v_2 w}.
\end{split}
\end{equation}

Concerning the neutral scalar sector, we find that the squared mass matrix for CP-even and CP-odd neutral scalars,
in the basis $\left( s_1,s_2,s_3 ,s_\phi,s'_1,s'_3 ,s_\sigma\right) $ and $\left( a_1,a_2,a_3 ,a_\phi,a'_1,a'_3 ,a_\sigma\right)$, are respectively given by:\\[-.2cm]
\begin{eqnarray}
M_{S} &=&\left( 
\begin{array}{cc}
M_{S}^{\left( 1\right) } & 0_{4\times 3} \\ 
0_{3\times 4} & M_{S}^{\left( 2\right) }%
\end{array}%
\right) , \\
M_{S}^{\left( 1\right) } &=&\left(
\begin{array}{cccc}
 2 \lambda _3 v_1^2+\frac{w v_2 \mu _t}{2 v_1} & v_1 v_2 \lambda _5-\frac{w \mu _t}{2} & w v_1 \lambda _6-\frac{v_2 \mu _t}{2} & \Lambda  v_1 \lambda _{12} \\
 v_1 v_2 \lambda _5-\frac{w \mu _t}{2} & 2 \lambda _1 v_2^2+\frac{w v_1 \mu _t}{2 v_2} & w v_2 \lambda _4-\frac{v_1 \mu _t}{2} & \Lambda  v_2 \lambda _{10} \\
 w v_1 \lambda _6-\frac{v_2 \mu _t}{2} & w v_2 \lambda _4-\frac{v_1 \mu _t}{2} & 2 \lambda _2 w^2+\frac{v_1 v_2 \mu _t}{2 w} & w \Lambda  \lambda _{11} \\
 \Lambda  v_1 \lambda _{12} & \Lambda  v_2 \lambda _{10} & w \Lambda  \lambda _{11} & 2 \Lambda ^2 \lambda _{16} \\
\end{array}
\right) , \\
M_{S}^{\left( 2\right) } &=&\frac{1}{2}\left(
\begin{array}{ccc}
 \frac{v_1 \left(w v_1 \lambda _9+v_2 \mu _t\right)}{w} & w v_1 \lambda _9+v_2 \mu _t & v_1 \left(\Lambda  \lambda _{19}+\mu _u\right) \\
 w v_1 \lambda _9+v_2 \mu _t & \frac{w \left(w v_1 \lambda _9+v_2 \mu _t\right)}{v_1} & w \left(\Lambda  \lambda _{19}+\mu _u\right) \\
 v_1 \left(\Lambda  \lambda _{19}+\mu _u\right) & w \left(\Lambda  \lambda _{19}+\mu _u\right) & \lambda _{14} w^2+2 \mu _5^2+v_2^2 \lambda _{13}+v_1^2 \lambda
   _{15}+\Lambda ^2 \lambda _{18}+2 \Lambda  \mu _s \\
\end{array}
\right) \nonumber,
\end{eqnarray}%
and
\begin{eqnarray}
M_{A} &=&\left( 
\begin{array}{cc}
M_{A}^{\left( 1\right) } & 0_{4\times 3} \\ 
0_{3\times 4} & M_{A}^{\left( 2\right) }%
\end{array}%
\right) , \\
M_{A}^{\left( 1\right) } &=&\frac{1}{2}\left(
\begin{array}{cccc}
 \frac{w v_2 \mu _t}{v_1} & w \mu _t & v_2 \mu _t & 0 \\
 w \mu _t & \frac{w v_1 \mu _t}{v_2} & v_1 \mu _t & 0 \\
 v_2 \mu _t & v_1 \mu _t & \frac{v_1 v_2 \mu _t}{w} & 0 \\
 0 & 0 & 0 & 0 \\
\end{array}
\right) , \\
M_{A}^{\left( 2\right) } &=&\frac{1}{2}\left(
\begin{array}{ccc}
 \frac{v_1 \left(w v_1 \lambda _9+v_2 \mu _t\right)}{w} & -w v_1 \lambda _9-v_2 \mu _t & v_1 \left(\Lambda  \lambda _{19}-\mu _u\right) \\
 -w v_1 \lambda _9-v_2 \mu _t & \frac{w \left(w v_1 \lambda _9+v_2 \mu _t\right)}{v_1} & w \left(\mu _u-\Lambda  \lambda _{19}\right) \\
 v_1 \left(\Lambda  \lambda _{19}-\mu _u\right) & w \left(\mu _u-\Lambda  \lambda _{19}\right) & \lambda _{14} w^2+2 \mu _5^2+v_2^2 \lambda _{13}+v_1^2 \lambda
   _{15}+\Lambda ^2 \lambda _{18}-2 \Lambda  \mu _s \\
\end{array}
\right)\nonumber .
\end{eqnarray}
The block $M_{S}^{\left( 1\right) } $ contains one small eigenvalue associated to the \sm Higgs field. Assuming the hierarchy $\Lambda,w,\mu_t\gg v_1, v_2$ the latter can be identified with
\begin{equation}
h\approx\frac{v_1s_1+v_2s_2}{\sqrt{v_1^2+v_2^2}},\qquad m^2_h =\mathcal{O}(v_{1,2}^2),
\end{equation}
and three heavy Higgs bosons, given as, 
\begin{equation}
\begin{split}
H_1&\approx\frac{v_2s_1-v_1s_2}{\sqrt{v_1^2+v_2^2}},\qquad m^2_{H_1}\approx \frac{\left(v_1^2+v_2^2\right) w \mu _t}{2 v_1 v_2},\\
H_2&\approx\cos\xi s_3-\sin\xi s_4,\qquad m^2_{H_2}\approx \lambda _{16} \Lambda ^2+\lambda _2 w^2-\sqrt{\lambda _{16}^2 \Lambda ^4+\lambda _2^2 w^4+\lambda _{11}^2 \Lambda ^2 w^2-2 \lambda _2 \lambda _{16} \Lambda ^2 w^2},\\
H_3&\approx\sin\xi s_3+\cos\xi s_4,\qquad m^2_{H_3}\approx \lambda _{16} \Lambda ^2+\lambda _2 w^2+\sqrt{\lambda _{16}^2 \Lambda ^4+\lambda _2^2 w^4+\lambda _{11}^2 \Lambda ^2 w^2-2 \lambda _2 \lambda _{16} \Lambda ^2 w^2}.
\end{split}
\end{equation}
The matrix $M_{A}^{\left( 1\right) } $ contains three Nambu-Goldstone bosons
\begin{equation}
G_1=\frac{v_1a_1-v_2a_2}{\sqrt{v_1^2+v_2^2}},\qquad
G_2=\frac{v_1a_1-w a_2}{\sqrt{v_1^2+w^2}},\qquad
G_3=a_\phi,\\
\end{equation}
related to the longitudinal
components of the $Z$, $Z^{\prime }$, $Z^{\prime \prime }$ gauge bosons, plus a heavy CP-odd massive state
\begin{equation}
A_1=\frac{v_2 w a_1+v_1 wa_2+v_1 v_2a_3}{\sqrt{(v_2 w)^2+(v_1 w)^2+(v_1 v_2)^2}},\qquad m^2_{A_1}= \frac{\mu _t \left(v_1^2 w^2+v_2^2 w^2+v_2^2 v_1^2\right)}{2 v_1 v_2 w}.
\end{equation}

The CP-even states $s'_1$, $s'_3$ and $s_\sigma$ mix according to the squared mass matrix $M_{S}^{\left( 2\right) } $, that can be diagonalized by the transformation
\begin{equation}
\left( 
\begin{array}{c}
\varphi_1 \\
\varphi_2 \\
G_4
\end{array}%
\right)=U^s \left( 
\begin{array}{c}
s'_1 \\
s'_3 \\
s_\sigma
\end{array}%
\right)=\left(
\begin{array}{ccc}
 \frac{v_1\cos\theta _s }{\sqrt{w^2+v_1^2}} & \frac{w \cos\theta _s}{\sqrt{w^2+v_1^2}} & \sin \theta _s\\
 -\frac{v_1\sin\theta _s }{\sqrt{w^2+v_1^2}} & -\frac{w \sin \theta _s}{\sqrt{w^2+v_1^2}} & \cos \theta _s \\
 \frac{w}{\sqrt{w^2+v_1^2}} & -\frac{v_1}{\sqrt{w^2+v_1^2}} & 0 \\
\end{array}
\right) \left( 
\begin{array}{c}
s'_1 \\
s'_3 \\
s_\sigma
\end{array}%
\right),
\end{equation}
with 
\begin{equation}
\tan2\theta_s=\frac{2 v_1 w \sqrt{v_1^2+w^2} \left(\lambda _{19} \Lambda +\mu _u\right)}{v_1 w \left(-2 \mu _5^2-\Lambda  \left(\lambda _{18} \Lambda +2 \mu _s\right)-\lambda
   _{13} v_2^2-\lambda _{15} v_1^2+\lambda _9 \left(v_1^2+w^2\right)-\lambda _{14} w^2\right)+v_2 \mu _t \left(v_1^2+w^2\right)},
\end{equation}
yielding two heavy physical real scalars $\varphi_1$ and $\varphi_2$ with squared masses
\begin{equation}
\begin{split}
&m_{\varphi_{1,2}}^2=\frac{1}{4 v_1 w}\Bigg\{v_1 w \left(\lambda _{18} \Lambda ^2+2 \mu _5^2+2 \Lambda  \mu _s+\lambda _{13} v_2^2+\lambda _{15} v_1^2+\lambda _9 \left(v_1^2+w^2\right)+\lambda _{14}
   w^2\right)+v_2 \mu _t \left(v_1^2+w^2\right)\\&\mp\mathcal{F}_s\Big\{\left(v_1 w \left(\lambda _{18} \Lambda ^2+2 \mu _5^2+2 \Lambda  \mu _s+\lambda _{13} v_2^2+\lambda _{15} v_1^2+\lambda _9
   \left(v_1^2+w^2\right)+\lambda _{14} w^2\right)+v_2 \mu _t \left(v_1^2+w^2\right)\right){}^2\\&-4 v_1 w \left(v_1^2+w^2\right) \big(v_2 \mu _t \left(\lambda _{18}
   \Lambda ^2+2 \mu _5^2+2 \Lambda  \mu _s+\lambda _{13} v_2^2+\lambda _{14} w^2\right)+v_1 w \big(\lambda _9 \left(\lambda _{18} \Lambda ^2+2 \mu _5^2+2 \Lambda 
   \mu _s+\lambda _{14} w^2\right)\\&-\left(\lambda _{19} \Lambda +\mu _u\right){}^2+\lambda _9 \lambda _{13} v_2^2\big)+\lambda _{15} v_2 v_1^2 \mu _t+\lambda _9
   \lambda _{15} v_1^3 w\big)\Big\}^{1/2}\Bigg\},\\
& \mathcal{F}_s=\mathrm{sgn}\left\{v_1 w \left(-2 \mu _5^2-\Lambda  \left(\lambda _{18} \Lambda +2 \mu _s\right)-\lambda
   _{13} v_2^2-\lambda _{15} v_1^2+\lambda _9 \left(v_1^2+w^2\right)-\lambda _{14} w^2\right)+v_2 \mu _t \left(v_1^2+w^2\right)\right\} , 
   \end{split}
 \end{equation}
and the Goldstone mode $G_4$. The CP-odd scalars $a'_1$, $a'_3$ and $a_\sigma$ have a similar fate, since the squared mass matrix $M_{A}^{\left( 2\right) }$ can be diagonalized by the transformation
\begin{equation}
\left( 
\begin{array}{c}
\widetilde{\varphi}_1 \\
\widetilde{\varphi}_2 \\
G_5
\end{array}%
\right)=U^a \left( 
\begin{array}{c}
a'_1 \\
a'_3 \\
a_\sigma
\end{array}%
\right)=\left(
\begin{array}{ccc}
 -\frac{v_1\cos \theta _a }{\sqrt{w^2+v_1^2}} & \frac{w \cos \theta _a}{\sqrt{w^2+v_1^2}} & \sin \theta _a \\
 \frac{v_1\sin \theta _a }{\sqrt{w^2+v_1^2}} & -\frac{w \sin \theta _a}{\sqrt{w^2+v_1^2}} & \cos \theta _a \\
 \frac{w}{\sqrt{w^2+v_1^2}} & \frac{v_1}{\sqrt{w^2+v_1^2}} & 0 \\
\end{array}
\right) \left( 
\begin{array}{c}
a'_1 \\
a'_3 \\
a_\sigma
\end{array}%
\right),
\end{equation}
with mixing angle
\begin{equation}
\tan2\theta_a=\frac{2 v_1 w \sqrt{v_1^2+w^2} \left(\mu _u-\lambda _{19} \Lambda \right)}{v_1 w \left(-\lambda _{18} \Lambda ^2-2 \mu _5^2+2 \Lambda  \mu _s-\lambda _{13}
   v_2^2-\lambda _{15} v_1^2+\lambda _9 \left(v_1^2+w^2\right)-\lambda _{14} w^2\right)+v_2 \mu _t \left(v_1^2+w^2\right)}.
\end{equation}
The real scalars $\widetilde{\varphi}_1$ and $\widetilde{\varphi}_2$ acquire squared masses
\begin{equation}
\begin{split}
&m_{\widetilde{\varphi}_{1,2}}^2=\frac{1}{4 v_1 w}\Bigg\{v_1 w \left(\lambda _{18} \Lambda ^2+2 \mu _5^2-2 \Lambda  \mu _s+\lambda _{13} v_2^2+\lambda _{15} v_1^2+\lambda _9 \left(v_1^2+w^2\right)+\lambda _{14}
   w^2\right)+v_2 \mu _t \left(v_1^2+w^2\right)\\&\mp\mathcal{F}_a\Big\{\left(v_1 w \left(\lambda _{18} \Lambda ^2+2 \mu _5^2-2 \Lambda  \mu _s+\lambda _{13} v_2^2+\lambda _{15} v_1^2+\lambda _9 \left(v_1^2+w^2\right)+\lambda _{14}
   w^2\right)+v_2 \mu _t \left(v_1^2+w^2\right)\right){}^2\\&-4 v_1 w \left(v_1^2+w^2\right) \big(v_2 \mu _t \left(\lambda _{18} \Lambda ^2+2 \mu _5^2-2 \Lambda  \mu _s+\lambda _{13} v_2^2+\lambda _{14} w^2\right)+v_1 w \big(\lambda _9 \left(\lambda _{18} \Lambda ^2+2 \mu _5^2-2 \Lambda  \mu _s+\lambda _{14} w^2\right)\\&-\left(\lambda _{19} \Lambda-\mu _u \right){}^2+\lambda _9 \lambda _{13} v_2^2\big)+\lambda _{15} v_2 v_1^2 \mu _t+\lambda _9
   \lambda _{15} v_1^3 w\big)\Big\}^{1/2}\Bigg\},\\
& \mathcal{F}_a=\mathrm{sgn}\left\{v_1 w \left(-\lambda _{18} \Lambda ^2-2 \mu _5^2+2 \Lambda  \mu _s-\lambda _{13}
   v_2^2-\lambda _{15} v_1^2+\lambda _9 \left(v_1^2+w^2\right)-\lambda _{14} w^2\right)+v_2 \mu _t \left(v_1^2+w^2\right)\right\} , 
   \end{split}
 \end{equation}
and the Goldstone boson $G_5$ combines with  $G_4$ into a neutral complex Goldstone associated with the non-Hermitian gauge boson $X^0$.
 Notice that in the limit $\mu_s,\mu_u\to 0$, one obtains a degenerate physical scalar spectrum $m_{\varphi_{1,2}}^2=m_{\widetilde{\varphi}_{1,2}}^2$.
 This degeneracy is broken in our model by the inclusion of the scalar singlet $\sigma$, a feature required to implement the scotogenic neutrino mass generation approach.

\section{Yukawa Sector}

\label{sec:Yukawa}

The Yukawa interactions and mass terms for fermions are given by 
\begin{equation}
\begin{split}
-\mathcal{L}_{\text{Yukawa}}=& y_{ab}^{e}\overline{l}_{aL}\rho
e_{bR}+y_{ab}^{N}\overline{l}_{aL}\chi N_{bR}+\frac{M_{Mab}}{2}\overline{N^{c}%
}_{aR}N_{bR} \\
& +y_{3a}^{u}\overline{q}_{3L}\eta u_{aR}+y_{ia}^{u}\overline{q}_{iL}\rho
^{\ast }u_{aR}+y^{U}\overline{q}_{3L}\chi U_{3R} \\
& +y_{3a}^{d}\overline{q}_{3L}\rho d_{aR}+y_{ia}^{d}\overline{q}_{iL}\eta
^{\ast }d_{aR}+y_{ij}^{D}\overline{q}_{iL}\chi ^{\ast }D_{jR}+\mathrm{h.c.}
\end{split}%
\end{equation}
After the spontaneous breakdown of the \3311 gauge symmetry, the Yukawa
interactions generate the following mass matrices for quarks :
\begin{equation}
M_{U}=\left( 
\begin{array}{cccc}
-y_{11}^{u}\frac{v_{2}}{\sqrt{2}} &- y_{12}^{u}\frac{v_{2}}{\sqrt{2}} & 
-y_{13}^{u}\frac{v_{2}}{\sqrt{2}} & 0 \\ 
-y_{21}^{u}\frac{v_{2}}{\sqrt{2}} & -y_{22}^{u}\frac{v_{2}}{\sqrt{2}} & 
-y_{23}^{u}\frac{v_{2}}{\sqrt{2}} & 0 \\ 
y_{31}^{u}\frac{v_{1}}{\sqrt{2}} & y_{32}^{u}\frac{v_{1}}{\sqrt{2}} & 
y_{33}^{u}\frac{v_{1}}{\sqrt{2}} & 0 \\ 
0 & 0 & 0 & y^{U}\frac{w}{\sqrt{2}}%
\end{array}%
\right) ,  \label{MU}
\end{equation}

\begin{equation}
M_{D}=\left( 
\begin{array}{ccccc}
y_{11}^{d}\frac{v_{1}}{\sqrt{2}} & y_{12}^{d}\frac{v_{1}}{\sqrt{2}} & 
y_{13}^{d}\frac{v_{1}}{\sqrt{2}} & 0 & 0 \\ 
y_{21}^{d}\frac{v_{1}}{\sqrt{2}} & y_{22}^{d}\frac{v_{1}}{\sqrt{2}} & 
y_{23}^{d}\frac{v_{1}}{\sqrt{2}} & 0 & 0 \\ 
y_{31}^{d}\frac{v_{2}}{\sqrt{2}} & y_{32}^{d}\frac{v_{2}}{\sqrt{2}} & 
y_{33}^{d}\frac{v_{2}}{\sqrt{2}} & 0 & 0 \\ 
0 & 0 & 0 & y_{12}^{D}\frac{w}{\sqrt{2}} & y_{21}^{D}\frac{w}{\sqrt{2}} \\ 
0 & 0 & 0 & y_{21}^{D}\frac{w}{\sqrt{2}} & y_{22}^{D}\frac{w}{\sqrt{2}}%
\end{array}%
\right).  \label{MD}
\end{equation}
Due to the $\mathrm{U(1)_{N}}$ symmetry assignments, there are no tree level mixing between the exotic and \sm (SM) quarks,
and therefore the Cabibbo-Kobayashi-Maskawa (CKM) matrix is unitary. As indicated by Eqs. (\ref{MU}) and (\ref{MD}),
both $\mathrm{SU(3)_{L}}$ scalar triplets $\eta $ and $\rho $ are needed to generate the up- and down-type SM quark masses, whereas the third triplet $\chi$, responsible
for the spontaneous breaking of the $\mathrm{SU(3)_{L}\otimes U\left( 1\right) _{X}}$ symmetry, produces the exotic quark masses.
 As shown from Eqs. (\ref{MU}) and (\ref{MD}) there is enough parametric freedom to successfully fit the experimental values of the SM quark masses and CKM parameters.
 
 It is worth mentioning that the non universal $\mathrm{U(1)_X}$ and $\mathrm{U(1)_N}$ charge assignments for the left handed quark fields give rise to flavour changing neutral processes (FCNC) mediated by the $Z^{\prime}$ and $Z^{\prime\prime}$ gauge bosons.
These contribute to the $K^0-\bar{K}^0$, $D^0-\bar{D}^0$ and $B^0_d-\bar{B}^0_d$ mass differences. As follows from Ref. \cite{Queiroz:2016gif}, the corrections to the mass differences of the $K^0$ and $D^0$ are quite small, whereas the corresponding corrections to the $B^0_d-\bar{B}^0_d$ mass difference can reach sizeable values not too far from the experimental sensitivity, for $Z^{\prime}$ and $Z^{\prime\prime}$ gauge boson masses in the few~TeV range.

Turning to the charged lepton sector, only the $\mathrm{SU(3)_{L}}$ scalar triplet $\rho $ contributes to the charged lepton mass matrix, given by
\begin{equation}
M_{l}=\left( 
\begin{array}{ccc}
y_{11}^{e} & y_{12}^{e} & y_{13}^{e} \\ 
y_{21}^{e} & y_{22}^{e} & y_{23}^{e} \\ 
y_{31}^{e} & y_{32}^{e} & y_{33}^{e}%
\end{array}%
\right) \frac{v_{2}}{\sqrt{2}}.  \label{Ml}
\end{equation}
Notice that, after SSB, the ``dark'' or $M_P$-odd fermions $N_L$ and $N_R$ mix through the mass matrix 
\begin{equation}
M_{N}=\left( 
\begin{array}{cc}
0& y^{N}\omega \\
(y^{N}\omega)^T&M_M
\end{array}%
\right),  \label{MN}
\end{equation}
in the basis $(N^c_L,N_R)$. Using the general method in Eq.(3.1) of~\cite{Schechter:1981cv} this matrix can be diagonalized perturbatively by a unitary transformation,
defining six physical Majorana states denoted by $S_{\alpha R}$ through
\begin{equation}
\left( 
\begin{array}{c}
N^c_L \\
N_R
\end{array}%
\right)=U S_R,  \label{SSt}
\end{equation}
such that $M'=U^TM_NU=\mathrm{diag}(M_\alpha)$. In the following discussion, only the lower blocks of the unitary matrix $U$ will be relevant.
We adopt the following notation for the relation between $N_R$ and $S_R$:
\begin{equation}
N_{aR}=U_{a\alpha} S_{\alpha R}. 
\end{equation}
For simplicity, we will assume that the entries of the matrix in Eq.(\ref{MN}) are real, and that the matrix $U$ becomes orthogonal.

\subsection{Neutrino masses}

Concerning the neutrino sector, the light active neutrino masses are produced by a radiative one-loop seesaw mechanism, thanks to the
remnant $M_{P}$ discrete symmetry preserved after the SSB of the $\mathrm{U(1)_{N}}$ gauge symmetry.
The key observation for the generation of light neutrino masses is the fact that the basic ingredients for a scotogenic mechanism are already present in the \3311 basic construction,
namely the fermionic dark singlets $N_R$ and the dark $\mathrm{SU(2)}$ scalar doublet, identified as the first two components of the $\chi$ triplet.
On a closer inspection, the Yukawa interaction that allows the existence of the diagram in Figure \ref{Scotogenicdiagram} splits into two pieces
\begin{equation}
y_{ab}^{N}\overline{l}_{aL}\chi N_{bR}=y_{ab}^{N}\left( 
\begin{array}{c c}
\overline{\nu}_a&\overline{e}_a
\end{array}%
\right)_L\left( 
\begin{array}{c}
\frac{s'_1+ia'_1}{\sqrt{2}}\\
\xi_2^{-}
\end{array}%
\right)N_{bR}+y_{ab}^{N}\overline{N}_{aL}\left(\frac{w+s_3+ia_3}{\sqrt{2}}\right) N_{bR}.
\end{equation}
The first term in the above relation is the necessary interaction between neutrinos and singlet fermions through an inert $\mathrm{SU(2)}$ scalar doublet, while the second term gives
rise to the Dirac mass blocks in Eq.(\ref{MN}). Thus, in the physical basis, the relevant terms for the generation of neutrino masses are
\begin{equation}
\begin{split}
-\mathcal{L}\supset&y_{ab}^{N}\overline{l}_{aL}\chi N_{bR}+\frac{M_{Mab}}{2}\overline{N^{c}%
}_{aR}N_{bR}
\\=&\frac{(y^NU)_{a\alpha}U^{s}_{i1}}{\sqrt{2}}\overline{l}_{aL}\varphi_{i} S_{\alpha R}+i\frac{(y^{N}U)_{a\alpha}U^{a}_{i1}}{\sqrt{2}}\overline{l}_{aL}\widetilde{\varphi}_{i} S_{\alpha R}\\
+&\frac{(y^NU)_{a\alpha}U^{s}_{31}}{\sqrt{2}}\overline{l}_{aL}G_4 S_{\alpha R}+i\frac{(y^{N}U)_{a\alpha}U^{a}_{31}}{\sqrt{2}}\overline{l}_{aL}G_5 S_{\alpha R}\\
&+\sum_{\alpha =1}^{6}\frac{M_{\alpha}}{2}\overline{S^{c}%
}_{\alpha R}S_{\alpha R}+\mathrm{h.c.}
\end{split}%
\end{equation}
Then, according to Fig. \ref{Scotogenicdiagram} the one loop level light active neutrino mass matrix is given by
\begin{equation}
\begin{split}
\left( M_{\nu }\right) _{ab}=&\sum_{\alpha =1}^{6}\sum_{i =1}^{2}\frac{(y^{N}U)_{a\alpha}(y^{N}U)_{b\alpha}M_\alpha}{%
16\pi ^{2}}\left[ (U^{s}_{i1})^2\frac{m_{\varphi_{i}}^2}{m_{\varphi_{i}}^{2}-M_{\alpha}^{2}}\ln \left( \frac{m_{\varphi_{i}}^{2}}{%
M_{\alpha}^{2}}\right) -(U^{a}_{i1})^2\frac{m_{\widetilde{\varphi}_i}^2}{m_{\widetilde{\varphi}_i}^{2}-M_{\alpha}^{2}}\ln \left( \frac{m_{\widetilde{\varphi}_i}^{2}}{%
M_{\alpha}^{2}}\right) \right] .
\end{split}
\end{equation}
where the mass splitting between the CP even and CP odd scalars running in the internal lines of the loop %
is generated from the $\frac{\mu _{s}}{\sqrt{2}} \phi^{\dagger} \sigma \sigma $ and $\frac{\mu _{u}}{\sqrt{2}}\left( \eta ^{\dagger}\chi \right) \sigma $ trilinear scalar interactions.
Thus, the tiny values of the light active neutrino masses can be attributed to the loop suppression, as well as to the smallness of the trilinear scalar couplings $\mu_{s}$ and $\mu_{u}$, which in turn produce a small mass splitting between the virtual CP even and CP odd scalars. We emphasize that the model under consideration has enough parametric freedom to successfully accommodate the experimental values of the neutrino mass squared differences, the leptonic mixing angles and the leptonic Dirac CP violating phase, as required by current neutrino experiments~\cite{deSalas:2017kay}.

\section{Dark Matter Phenomenology}
\label{sec:DM} 
In this section we will discuss the implications of our model in the Dark matter sector. Due to the residual matter-parity symmetry, our model has stable scalar and fermionic dark matter candidates. The scalar dark matter candidate will be the lightest of the physical scalar fields charged under the conserved matter-parity symmetry, whereas the lightest electroweak singlet $N_R$ is the fermionic candidate. In what follows we discuss separately the scenarios where the dark matter candidate is either scalar or fermion.\vfill

\subsection{Scalar Dark matter candidates}

In order to show the viability of our model as a theory of dark matter, we analyze a simplified scenario in which all the non-SM fields are heavy and decouple, except for the complex scalar $\varphi_2$, identified as our scalar dark matter candidate.
Working in the limit of a small mixing angle $|\theta_s|<<1$, the scalar field $\varphi_2$ is mostly composed by the electroweak singlet $\sigma$, yielding a small coupling between the complex dark matter candidate and the $Z$-boson, consistent with direct detection experiments. In this approximation, our dark matter candidate mainly annihilates into $hh$ via the Higgs portal quartic scalar interaction $\lambda_{\mathrm{eff}} h^{2}\left(\varphi_2\right) ^{2}$. For adequate values of the scalar DM mass and the effective coupling  $\lambda_{\mathrm{eff}}$
the experimental relic density value can be accomodated, see~\cite{CarcamoHernandez:2016pdu,Bernal:2017xat,CarcamoHernandez:2017kra,Long:2018dun,Kang:2019sab}. 
Concerning dark matter direct detection prospects, the scalar DM candidate would scatter off a nuclear target in a detector via Higgs boson exchange in the $t$-channel,
thus constraining the $\lambda_{\mathrm{eff}}$ coupling.
A simple analytical estimate can be performed in the case where $m_{\varphi_2}^{2}\gg v^{2}$, with $v=246$ GeV.
Neglecting the annihilation channel of the scalar DM candidate into neutrino-antineutrino pairs as in ~\cite{Bernal:2017xat}, the freeze-out of heavy scalar DM particle is largely
dominated by the annihilation into Higgs bosons, and the corresponding thermally averaged cross section can be estimated as 
\begin{equation}
<\sigma v>\simeq \frac{\lambda_{\mathrm{eff}}^{2}}{128\pi m_{\varphi_2}^{2}},
\end{equation}%
which results in a DM relic abundance 
\begin{equation}
\frac{\Omega _{DM}h^{2}}{0.12}=\frac{0.1\,\mathrm{pb}}{0.12<\sigma v>}\simeq \left( 
\frac{1}{\lambda_{\mathrm{eff}}}\right) ^{2}\left( \frac{m_{\varphi_2}}{1.1\,\mathrm{TeV}}\right) ^{2},
\label{RDestimate}
\end{equation}%
showing that the our model is able to reproduce the relic density observed value \cite{Aghanim:2018eyx}
\begin{equation}
\Omega _{DM}h^{2}=0.120,  \label{DMrelicdensity}
\end{equation}%
for natural values of the effective quartic coupling $\lambda_{\mathrm{eff}}$ and a dark matter mass of $\mathcal{O}(1\,\mathrm{TeV})$. \\[-.2cm]

\begin{figure}[!h]
\begin{center}
\includegraphics[width=0.6\textwidth]{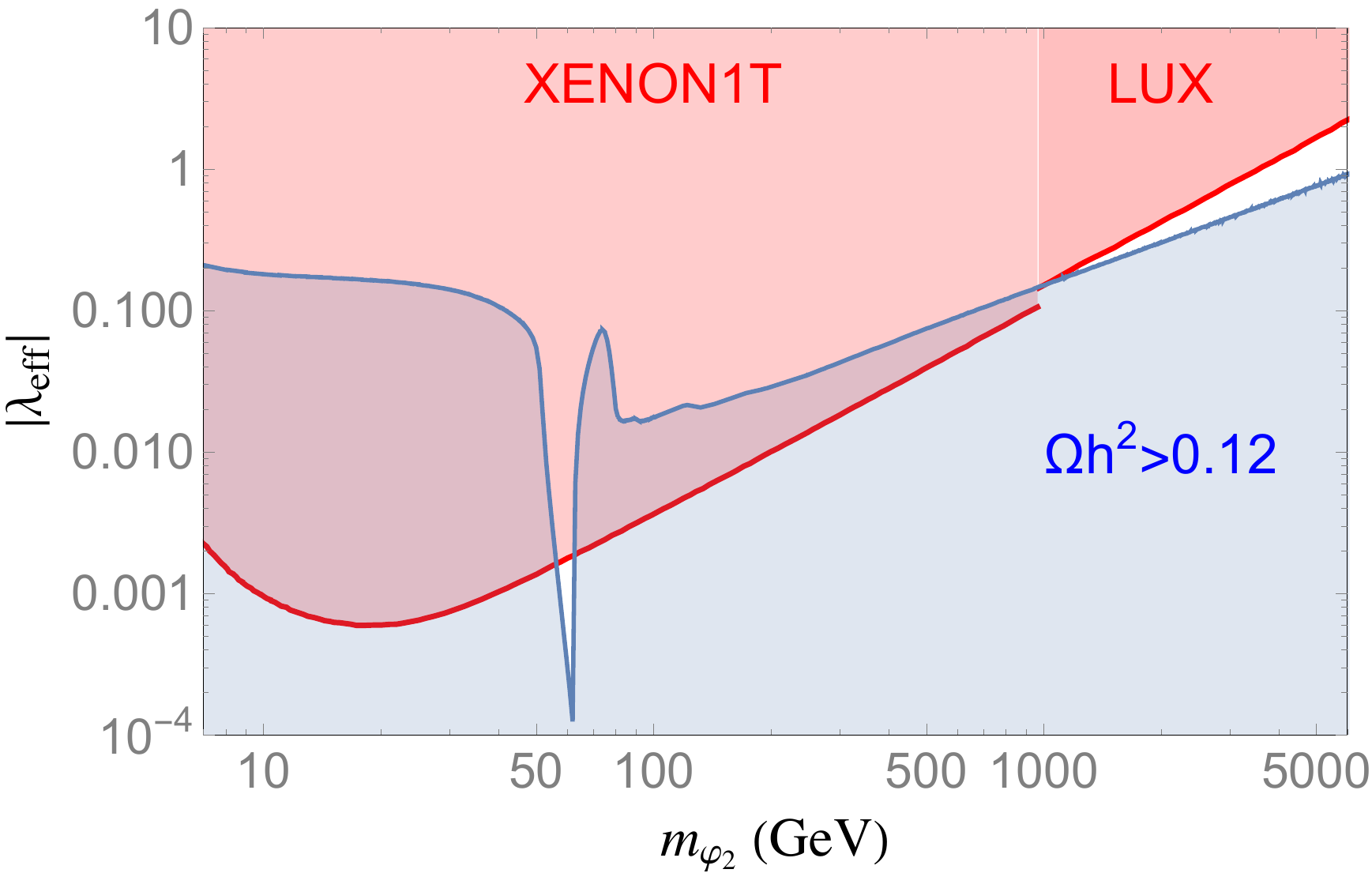}
\caption{The direct detection and relic abundance constraints on the dark matter mass $m_{\varphi_2}$ for the simplified scenario described in the text. The red shaded regions are ruled out by direct detection experiments, XENON1T \cite{Aprile:2018dbl} (below $1\,\mathrm{TeV}$) and LUX \cite{Akerib:2016vxi} (above $1\,\mathrm{TeV}$). The blue shaded region is not compatible with the measured dark matter relic abundance \cite{Aghanim:2018eyx}.}
\label{DM}
\end{center}
\end{figure}
A more detailed analysis can be performed numerically in the same simplified scenario where the single parameter $\lambda_{\mathrm{eff}}$, responsible for both the relic abundance as well as the
  direct detection cross section.
In Figure \ref{DM} we show the restrictions on the $m_{\varphi_2}-\lambda_{\mathrm{eff}}$ plane obtained by requiring the correct dark matter relic abundance, while simultaneously imposing the dark matter direct detection constraints from  XENON1T \cite{Aprile:2018dbl} and LUX \cite{Akerib:2016vxi}.
  Within our approximations the dark matter candidate $m_{\varphi_2}$ yields viable relic densities in only two distinct mass regions. The first one is near half the Higgs mass, where resonant annihilation of dark matter into the Higgs boson takes place,  allowing the relic density constraints to be satisfied for very small values of $\lambda_{\mathrm{eff}}$, well below the current direct detection bounds.
  The second region coincides with the previously discussed analytical estimate, starting at around 1 TeV, where the direct detection constraints on the coupling $\lambda_{\mathrm{eff}}$ are weak.
  This result illustrates that, even within this constrained scenario, our model can provide a viable scalar dark matter candidate.

We stress that Figure \ref{DM} assumes a simplified scenario in which only the Higgs portal is available for a single scalar. This needs not be the case.
  In our model, the allowed parameter space can be considerably richer due to re-scattering effects coming from the other available scalar dark fields \cite{Kakizaki:2016dza} and the from inclusion of Majorana fermions, like $N_R$, providing new channels for dark matter annihilation.
  Similarly, more parameter combinations become available when the additional neutral vector boson portals are active in mediating dark matter annihilation processes, instead of simply decoupled, as assumed in the above example. 

\subsection{Fermionic Dark matter candidates}

  Concerning the case of a fermionic DM candidate, we can estimate the DM relic density in a simplified scenario where only one candidate is light and the remaining non-SM fields are heavy and decoupled. In the following analysis we assume that the light DM candidate is $N_{1R}$ and that the mixing in Eq.(\ref{SSt}) is small enough to consider $N_{1R}$ as an approximate mass eigenstate.
From the Yukawa interaction $y_{ab}^{N}\overline{l}_{aL}\chi N_{bR}$ it follows that the DM candidate can annihilate into $\varphi_{1,2}\varphi _{1,2}$ and $\widetilde{\varphi}_{1,2}\widetilde{\varphi}_{1,2}$ via the $t$ channel exchange of the SM neutrinos $\nu_{iL}$ ($i=1,2,3$).
Furthermore, the fermionic DM candidate can also annihilate into a pair of SM neutrinos via the $t$ channel exchange of $\varphi_{1,2}$\ and $\widetilde{\varphi}_{1,2}$.
The resulting Dark matter relic density will depend on the neutrino Yukawa coupling, on the fermionic DM candidate mass $m_{N_{1R}}$,
as well as on the parameters $m_{\varphi _{1,2}}^{2}$ and $m_{\widetilde{\varphi}_{1,2}}^{2}$.

In a scenario where $m_{N{_{1R}}}^{2}{<}m_{\varphi _{1,2}}^{2}\sim m_{\widetilde{\varphi}_{1,2}}^{2}$, and the annihilation channel $N_{1R}N_{1R}\rightarrow \nu _{i}\nu _{i}$ ($i=1,2,3$), following \cite{Bernal:2017xat}
one can estimate the corresponding thermally averaged cross section as 
\begin{equation}
<\sigma v>\simeq \frac{9\left( y_{11}^{N}\right)
^{4}m_{N_{1R}}^{2}}{32\pi m_{\varphi _{1}}^{4}},
\end{equation}
which implies that the DM relic abundance takes the form: 
\begin{equation}
\frac{\Omega _{DM}h^{2}}{0.12}=\frac{0.1\,\mathrm{pb}}{0.12<\sigma v>}\simeq \left( 
\frac{1}{y_{11}^{N}}\right) ^{4}\left( \frac{400\,\mathrm{GeV}}{%
m_{N_{1R}}}\right) ^{2}\left( \frac{m_{\varphi _{1}}}{1.6\,\mathrm{TeV}}\right) ^{4},
\end{equation}%
indicating that in the scenario of a fermionic DM candidate, the observed value (\ref{DMrelicdensity}) of the Dark matter relic density can be correctly reproduced for
reasonable values of the Yukawa coupling, fermionic dark matter candidate mass and $m_{\varphi_{1}}$. Again, we stress that this is an oversimplified approximation.

\section{Discussion}
\label{sec:Discussion}

In this letter we have explored the idea that dark matter stability results from a residual matter-parity symmetry that survives the spontaneous
breaking of an extended gauge symmetry.
For the latter we have taken the \3311 symmetry, proposing the simplest scotogenic dark matter completion of the original SVS theory~\cite{Singer:1980sw}.
In our new construction the ``dark sector'' particles are clearly identified with states already present in the original picture.
The only new state added is a new singlet scalar in order to break the degeneracy of the neutral scalars.
The latter is also needed to close the scotogenic Majorana neutrino mass loop.
The theory provides a natural origin for the scalar dark doublet introduced \textit{ad hoc} in other dark matter constructions, such as the inert Higgs dark matter scenarios.
Here the latter is simply identified with the electroweak doublet part one of the triplet Higgs scalars required for adequate breaking of the extended $\mathrm{SU(3)_L}$ gauge symmetry.
Assuming this scalar to be the lightest ``odd particle'' it will be dark matter, with its stability naturally ensured by the residual matter-parity gauge symmetry.
This gives an elegant scotogenic realization of inert doublet scenarios of dark matter.
 We have also given simple estimates demonstrating the viability of our dark matter scenario.
 More extensive, dedicated studies along the lines of Refs.~\cite{Diaz:2015pyv,Arbelaez:2016mhg,Garcia-Cely:2015khw,Rojas-Abatte:2017hqm,Dutta:2017lny,Nomura:2017kih,Gao:2018xld,CarcamoHernandez:2019cbd,Bhattacharya:2019fgs,Han:2019lux} would be desirable and worth performing.

\acknowledgements 

Work supported by Spanish grants FPA2017-85216-P (AEI/FEDER, UE), PROMETEO/2018/165 (Generalitat Valenciana)
and the Spanish Red Consolider MultiDark FPA2017-90566-REDC.
CAV-A is supported by the Mexican Catedras CONACYT project 749 and SNI 58928.
A.E.C.H. received funding from Fondecyt (Chile), Grant No.~1170803. Numerical work performed in GuaCAL (Guanajuato Computational Astroparticle Lab). The relic abundance and
direct detection constraints are calculated using the MicroOmegas package \cite{Belanger:2018ccd}.

\bibliographystyle{utphys}
\bibliography{bibliography}

\end{document}